\documentclass{aa}

\usepackage{color}
\usepackage{graphicx}
\usepackage[]{subfig}
\usepackage{txfonts}
\usepackage[normalem]{ulem} 
\usepackage{soul}

\newcommand{\msun}{M$_\odot$}

\newcommand{\kms}{km s$^{-1}$}

\begin{document}

   \title{Lack of high-mass pre-stellar cores in the starless MDCs of NGC6334}


   \author{F. Louvet
          \inst{1}
          \and
         S. Neupane\inst{1}
         \and
        G. Garay\inst{1}
         \and
        D. Russeil\inst{2}
         \and
        A. Zavagno\inst{2}
         \and
        A. Guzman\inst{1}
         \and
        L. Gomez\inst{4}
         \and
        L. Bronfman\inst{1}
          \and
        T. Nony\inst{3}
          }

   \institute{Universidad de Chile, Camino el Observatorio 1515, Las Condes, Santiago de Chile
              \email{flouvet@das.uchile.cl}
\and
             Aix Marseille Univ., CNRS, LAM, Laboratoire d'Astrophysique de Marseille, Marseille, France
\and
            Univ. Grenoble Alpes, CNRS, IPAG, F-38000 Grenoble, France
\and
             Joint ALMA Observatory, Alonso de Córdova 3107, Vitacura, Santiago, Chile
             }

   \date{Received November 12, 2017; accepted December 1, 2017}

 
  \abstract
   {The formation of high-mass stars remains unknown in many aspects. Two families of models compete to explain the formation of high-mass stars. On the one hand, quasi-static models predict the existence of high-mass pre-stellar cores sustained by a high degree of turbulence. On the other hand competitive accretion models predict that high-mass proto-stellar cores evolve from low/intermediate mass proto-stellar cores in dynamic environments.}
   {The aim of the present work is to bring observational constraints at the scale of high-mass cores ($\sim$0.03 pc).}
   {We targeted with ALMA and MOPRA a sample of 9 starless massive dense cores (MDCs) discovered in a recent Herschel/HOBYS study. Their mass and size ($\sim$110 \msun~and $r$=0.1 pc, respectively) are similar to the initial conditions used in the quasi-static family of models explaining for the formation of high-mass stars. We present ALMA 1.4mm continuum observations that resolve the Jeans length ($\lambda_{\rm Jeans}\sim$0.03 pc) and that are sensitive to the Jeans mass (M$_{\rm Jeans}\sim$0.65~\msun) in the 9 starless MDCs, together with ALMA-$^{12}$CO(2-1) emission line observations. We also present HCO$^{+}$(1-0), H$^{13}$CO$^{+}$(1-0) and N$_2$H$^+$(1-0) molecular lines from the MOPRA telescope for 8 of the 9 MDCs.}	
   {The 9 starless MDCs have the mass reservoir to form high-mass stars according to the criteria by \cite{baldeschi17}. Three of the starless MDCs are subvirialized with $\alpha_{\rm vir}\sim$0.35, and 4 MDCs show sign of collapse from their molecular emission lines. ALMA observations show very little fragmentation within the MDCs. Only two of the starless MDCs host compact continuum sources, whose fluxes correspond to $<3$~\msun~ fragments. Therefore the mass reservoir of the MDCs has not yet been accreted onto compact objects, and most of the emission is filtered out by the interferometer.}
   {These observations do not support the quasi-static models for high-mass star formation since no high-mass pre-stellar core is found in NGC6334. The competitive accretion models, on the other hand, predict a level of fragmentation much higher than what we observe.}


   \maketitle
%

\section{Introduction}

Two main theoretical scenarios have been proposed to explain the formation of high-mass stars: \textit{(1)} monolithic collapse of a massive dense core (MDC, $\sim$100 \msun~for a radius of $\sim$0.1 pc) supported by supersonic turbulent pressure \citep[e.g.,][]{mckee-tan02,krumholz07} or \textit{(2)} competitive accretion in a proto-cluster environment through Bondi–Hoyle accretion \citep[e.g.,][]{bonnell04, bonnell-bate06}. These two scenarios lead to distinct characteristics for the initial stages of high-mass star formation. The first family of models suppose the existence of starless MDCs that are supported by high-degrees of micro-turbulence \citep[$\sigma\sim$1.7-2 \kms, ][]{krumholz07}. The MDCs contract quasi-statically to become high-mass pre-stellar cores ($\sim$30 \msun~for a radius of $\sim$0.03 pc) before becoming proto-stellar cores. Here, ``core''  names a gaseous structure of $\sim$0.03 pc \citep{bontemps10,zhang14, palau15} that will collapse to form a single star or a small N-tuple binary system. Hence, quasi-static models predict the existence of one, or a few, high-mass pre-stellar cores in the starless MDCs.

\smallskip 

In the second family of models, high-mass pre-stellar cores never develop. The starless MDCs fragment into a cluster of low-mass cores with initial masses of the order of the Jeans mass. The protostars inside these cores start accreting gas from their envelope, unimpeded by the presence of other protostars. In a second step, the gas is accreted from regions outside of their envelopes and therefore from region that can contribute mass to more than one protostar. This accretion accounts for the bulk of the mass of the more massive stars. The latter form favourably at the centre of the MDCs, where the gas density is the highest. 

A few high-resolution studies have been performed with (sub)millimeter interferometers with the aim of identifying pre-stellar cores and protostars within massive dense cores \citep[MDCs, see ][]{rathborne07, swift09, zhang09, busquet10, bontemps10, pillai11, wang11, zhang-wang11, beuther13, lee13, tan13, wang14, louvet14, cyganowski14, ana14, fontani16, kong17, palau18, nony18}. 
 These attempts revealed either that they were proto-stellar in nature when observed at high resolution \cite[e.g.][]{ana13}, or filled with low-mass pre-stellar cores \citep[e.g.][]{tan13, kong17}. In total, only five high-mass pre-stellar core candidates have been reported:
\begin{itemize}
\item[$\bullet$] The pre-stellar candidate CygX-N53-MM2 ($\sim$25 \msun~within 0.025 pc) of \cite{ana14}; however, owing to the confusion with the neighbour CygX-N53-MM1 it is hard to exclude that CygX-N53-MM2 is driving outflows.
\item[$\bullet$] The pre-stellar candidate G11.92-0.61-MM2 ($>$30\msun~within 0.01 pc) of \cite{cyganowski14} but the lack of (sub)millimeter molecular line emissions casts doubt about its belonging to the Milky Way.  
\item[$\bullet$] The pre-stellar candidate G11.11-P6-SMA1 ($\sim$30 \msun~within 0.02 pc) of \cite{wang14}. This source seems deprived of outflows but a spectral survey of this target is necessary to determine if it hosts a hot core.
\item[$\bullet$] The source G028CA9 ($\sim$70 \msun~within 0.04 pc) of \cite{kong17} but this source still lacks of shock tracer analysis to address if it is driving outflows. Also, according to the authors, it shows complex kinematics potentially indicative of two merging structures.
\item[$\bullet$] The source W43-MM1-\#6 ($\sim$55 \msun~within 1300 au) of \cite{nony18}. The target seems deprived of outflows and its spectral survey did not reveal strong complex organic molecule emissions. Nevertheless the study could not rule out the existence of an unresolved outflow and/or a small and embedded hot core.
\end{itemize}
Therefore although many regions have been studied very few high-mass pre-stellar core candidates have been reported which proves that, if they exist, they are very elusive. Here, we present a study of 9 starless MDCs located in NGC6334.

NGC6334 is a giant molecular cloud complex ($\sim$7$\times10^5$ \msun, see e.g. \citealt{russeil12}) of the Milky Way. It belongs to the Sagittarius-Carina arm, lying at 1.75 kpc from the sun \citep{matthews08}. It is a very active high-mass star-forming region hosting more than 2000 young stellar objects identified with \textit{Spitzer} \citep{willis13}, numerous H{\,\sc{ii}} regions, maser sources, and molecular outflows \citep[see][]{loughran86,persi-tapia08,carral02}. As part of the \textit{Herschel}/HOBYS project, \cite{tige17} analyzed the properties of the MDCs forming in NCG6334, based on \textit{Herschel}-PACS\&SPIRE, JCMT-SCUBA2, APEX-LABOCA, SEST-SIMBA, \textit{Spitzer}-IRAC\&MIPS, \textit{WISE}, and \textit{MSX} data. Among other results, they found and characterized 16 candidate starless massive dense cores that remained undetected at 70 $\mu$m, have a mass $>$70 \msun, and a deconvolved radius of $\sim$0.1 pc. Thirteen of these MDCs were observed with ALMA (2015.1.00850.S, PI: Louvet, F.). In this article we present 9 of these MDCs (see Table~\ref{t:obs}), as the other 4 contain protostars and will be presented separately (Louvet et al., in prep). We complement our data-set with MOPRA observations to analyse the large-scale ($\sim$0.1 pc) features of the MDCs. The article is organized such as Section \ref{s:obs} describes the observational data used in the present work, Sects.~\ref{ss:ls}-\ref{ss:infall} present the large-scale feature of the MDCs at $\sim$0.1 pc scales, Sect.~\ref{ss:ss} presents the structure of the MDCs down to $\sim$0.006 pc scales with ALMA. Section~\ref{s:discu} discusses our results with respect to the numerical models of high-mass star formation (\ref{ss:models}) and with respect to recent interferometric studies (\ref{ss:dobs}). Finally, Sect. \ref{s:conclu} exposes our conclusions.

\section{Observations}
\label{s:obs}

\subsection{ALMA observations}
\label{ss:obs-ss}

ALMA observations were carried out in Cycle 3 in Band 6 using 39 antennas with baselines ranging from 15 m to 459 m, that gives access to an angular resolution of $\sim$0.74$^{\prime\prime}$ and a filtering scale of 11.6$^{\prime\prime}$ at 1.4 mm. In this article, we only present continuum emission observations and $^{12}$CO(2-1) emission line observations. The continuum is obtained from a bandwidth spectral window of 1.875 GHz centred at 214.5 GHz (or 1.40~mm) divided in 128 channels of 31.25 MHz each. The $^{12}$CO(2-1) molecular line emission is obtained from a spectral windows centred on the $^{12}$CO(2-1) transition at 230.53 GHz with a high spectral resolution of 0.485 MHz (or 0.63 \kms). Each of our 9 sources was observed for $\sim$1.4 min in total. The system temperature has a mean value of 91 K. 
The data was pipeline-calibrated using standard procedures of CASA 4.5.2. The continuum was imaged concatenating together the 128 channels of the spectral window dedicated to the continuum emission, which is a valid approach for pre-stellar structures for which no complex organic molecule emission lines are expected \citep{motte18}. The $^{12}$CO(2-1) emission line was slightly smoothed to a spectral resolution of 0.7 \kms. Both the continuum and $^{12}$CO line emissions were imaged with the \textit{Clark} method using the weighting \emph{Briggs} with a robust parameter of 0.5 \footnote{Details on the method can be found in the CASA Task reference manual at the following location: https://casa.nrao.edu/docs/taskref/tclean-task.html}. The mean achieved root mean square (rms) noise level is of $\sim$0.34 mJy/beam in continuum and of 16.6 mJy/beam per channel in $^{12}$CO(2-1), at a resolution of 0.7 \kms. The Table~\ref{t:anex} details the rms for each source.

\subsection{MOPRA observations}
\label{ss:obs-ls}

These observations were performed in September 2008 with the ATNF MOPRA 22-m telescope with the 3mm receiver and the MOPRA spectrometer (MOPS) 
in the ``zoom mode'' that allows to observe simultaneously up to 16 different frequencies, among them the HCO$^+$(1-0) at 89.18 GHz, the H$^{13}$CO$^+$(1-0) at 86.75 GHz, and the N$_2$H$^+$(1-0) at 93.17 GHz emission lines.
Small maps (between $3^\prime\times 3^\prime$ to $5^\prime\times 5^\prime$) were observed in the direction of the 40 dense cores of \cite{russeil10}. These maps cover 7 of our 9 MDCs.
These maps were done in the OTF mode and the data reduction was performed with Livedata and Gridzilla tools\footnote{see http://www.atnf.csiro.au/computing/software/index.html} producing
data-cubes with a spectral and spatial resolution of 0.11 \kms and 30$^{\prime\prime}$ respectively. The MOPRA data-set is complemented with observations from the MALT90 survey \citep{foster13} to cover one additional MDC. To sum up, we gather a coherent sample of spectral lines from the MOPRA telescope for all our 9 MDCs but the MDC \#44.

\section{Analysis}
\label{s:ana}

\subsection{Large-scale view of the MDCs}
\label{ss:ls}

\begin{table*}
\caption{Physical parameters of the MDCs}
\label{t:obs}
\centerline{
\addtolength{\tabcolsep}{+1pt}
\begin{tabular}{cclccccccccccc}
\hline
\hline                                                                                                                                                                                  
MDC                   & Radius    & Mass           &${\rm v_{N_2H^+}}$ & $\sigma_{\rm N_2H^+}$   & ${\rm v_{HCO^+}}$  & ${\rm v_{H^{13}CO^+}}$ & $\delta$V      & M$_{vir}$         & $\alpha_{\rm vir}$     & M$_{thr}$       & M$_{\rm Jeans}$   & $\lambda_{\rm Jeans}$    & Bfield    \\
                      & [pc]      & [M$_\odot$]    & [km/s]            & [km/s]                  & [km/s]             & [km/s]                 &   -            & [M$_\odot$]       & -                      & [M$_\odot$]     & [M$_\odot$]       & [mpc]                    &  [mG]     \\
(1)                   & (2)       & (3)            & (4)               & (5)                     & (6)                & (7)                    &  (8)           & (9)               & (10)                   & (11)            & (12)              & (13)                     & (14)      \\
\hline                                                                                                                                                                                                                                                                                                          
NGC6334-MDC\#11       & 0.140     & 170            &  -                & -                       & -1.8               & -                      &  -             &   -               &    -                   &  63.7           & 0.85              & 46                       & 0.7       \\ 
NGC6334-MDC\#16       & 0.075     & 150            &  -4.0             & 1.00                    & -5.6               & -3.7                   &  -1.6          &  60               & 0.40                   &  27.8           & 0.21              & 15                       & 2.2       \\ 
NGC6334-MDC\#17       & 0.050     & 150            &  -7.3             & 1.72                    & -7.8               & -                      &  -0.3          & 120               & 0.80                   &  16.2           & 0.25              & 11                       & -         \\ 
NGC6334-MDC\#21       & 0.135     & 130            &  -4.9             & 0.99                    & -4.6               & -4.7                   &  +0.3          & 107               & 0.80                   &  85.7           & 1.31              & 55                       & -         \\ 
NGC6334-MDC\#27       & 0.070     & 105            &  -5.4             & 0.77                    & -5.5               & -5.4                   &  -0.1          &  34               & 0.30                   &  25.3           & 0.42              & 21                       & 1.9       \\ 
NGC6334-MDC\#26       & 0.040     & 100            &  -5.3             & 1.60                    & -6.3               & -4.8                   &  -0.6          &  83               & 0.80                   &  12.0           & 0.28              & 10                       & -         \\ 
NGC6334-MDC\#35       & 0.105     &  90            &  -1.9             & 0.65                    & -2.0               & -                      &  -0.2          &  36               & 0.35                   &  43.4           & 0.93              & 42                       & 0.7       \\ 
NGC6334-MDC\#44       & 0.150     &  77            &  -                & -                       & -                  & -                      &  -             &   -               &  -                     &  69.8           & 1.15              & 68                       & -         \\ 
NGC6334-MDC\#45       & 0.055     &  76            &  -8.9             & 1.12                    & -9.2               & -8.9                   &  -0.3          &  56               & 0.75                   &  16.4           & 0.34              & 17                       & 0.4       \\ 
\hline                                                                                                                                                                                                                                                                                                                                           
Mean                  & 0.090     & 116            &                   & 1.12                    &                    &                        &                &  71               & 0.60                   &  40.0           & 0.64              & 32                       & 1.2       \\
\hline                                                                                                                                                                                                                                     
\end{tabular}}
\end{table*}

\begin{figure}[h!]
\centerline{
\includegraphics[trim = 0cm 0cm 0cm 0cm, width=0.5\textwidth]{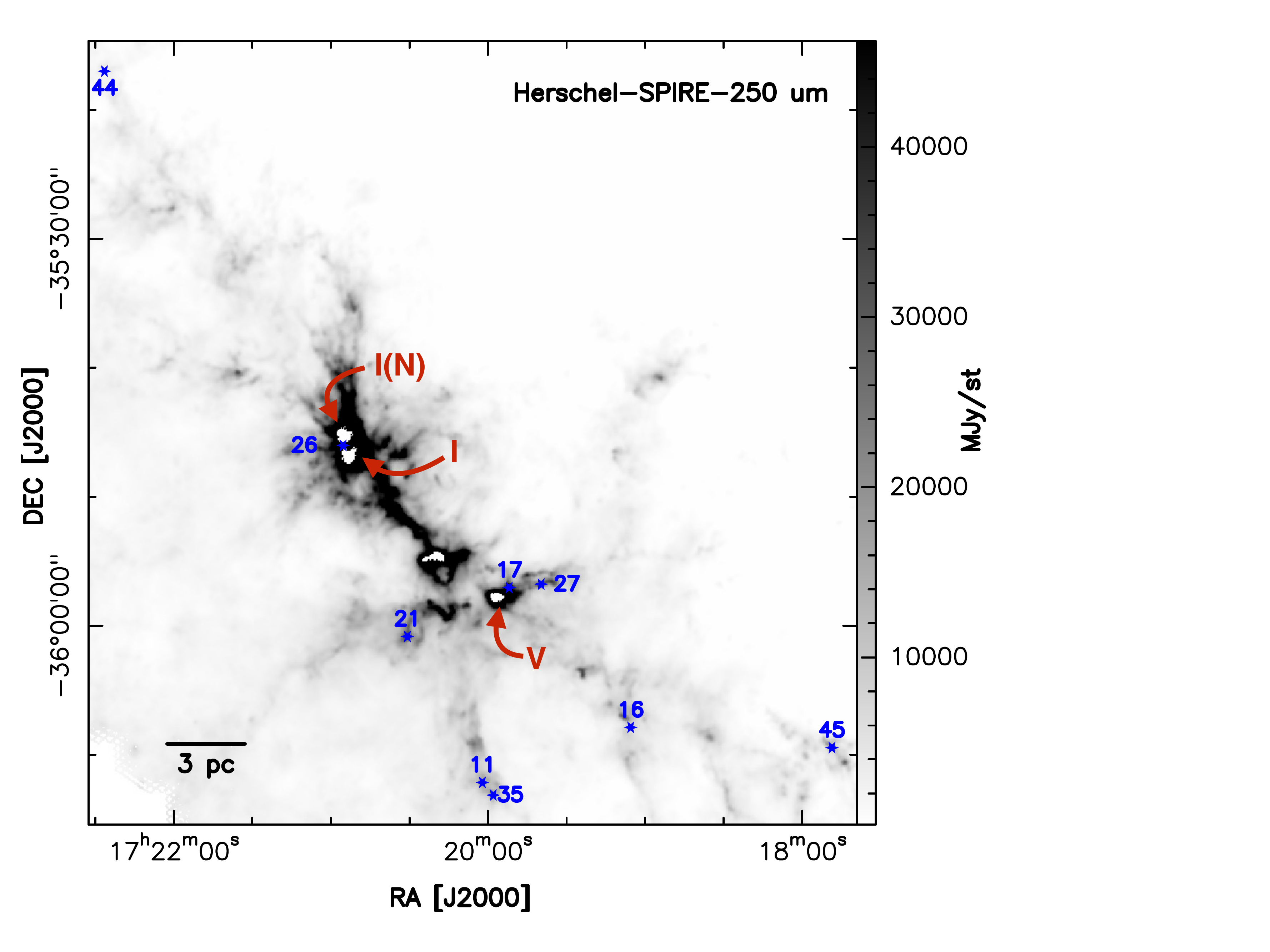}}
\caption{\textit{Herschel}-SPIRE-250 $\mu$m emission of the NGC6334 region. The 9 MDCs of the present work are located by the blue stars together with their number, following the nomenclature by \cite{tige17}. Red arrows and Roman numerals (I, V) denote bright far-infrared sources from \cite{kraemer-jackson99}. Credits: ESA, \textit{Herschel} Science archive.}
\label{f:ls}
\end{figure}

\begin{figure*}[h!]
\centerline{
\includegraphics[trim = 0cm 0cm 13cm 0cm , width=0.8\textwidth]{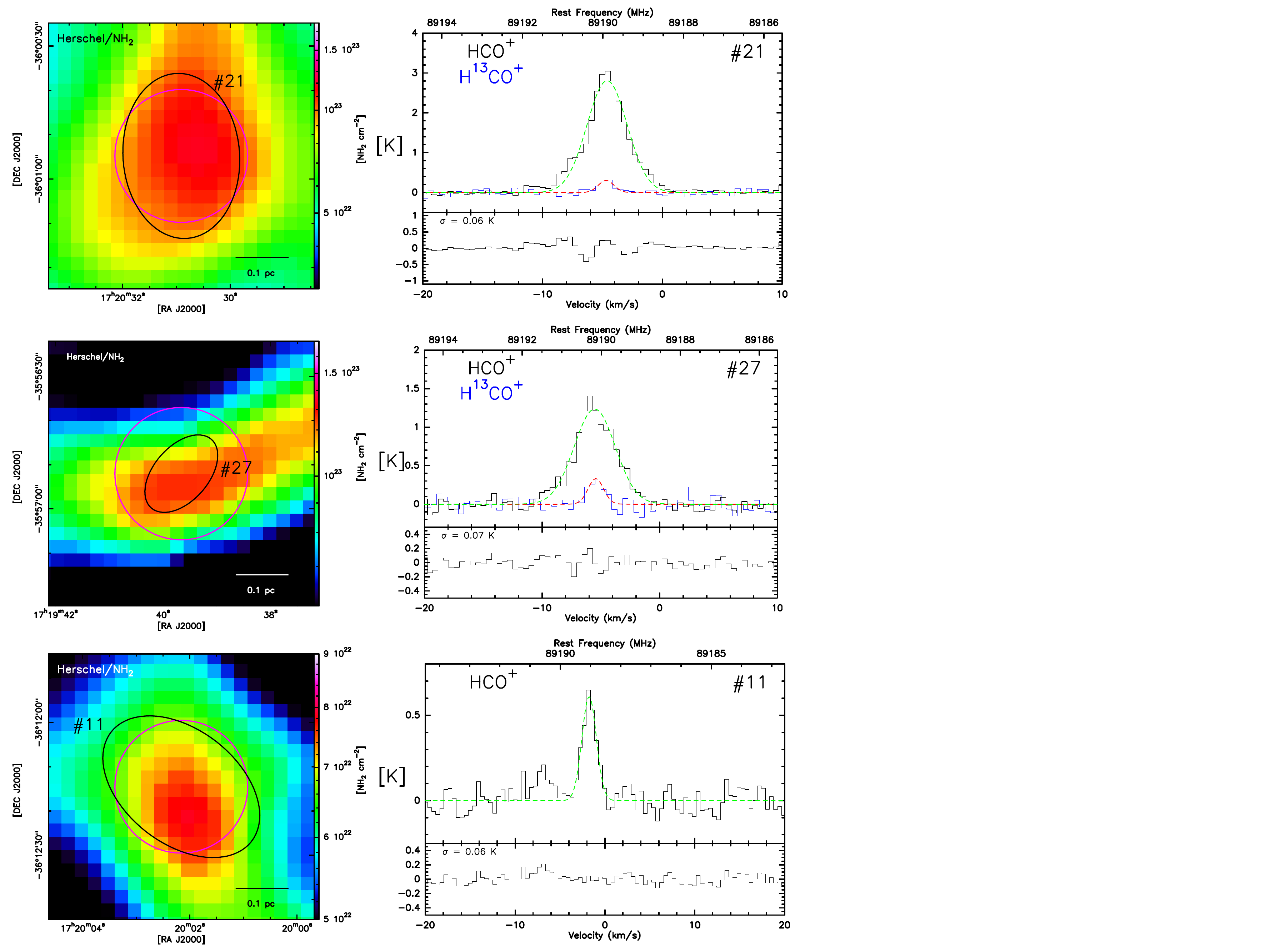}}
\caption{\textbf{Left:} \textit{Herschel}/column density maps toward the 9 MDCs studied here. The black ellipse indicates the position and size of the MDC as extracted by \cite{tige17}, and the magenta circle shows the beam of the MOPRA telescope. \textbf{Right:} HCO$^{+}$(1--0) and H$^{13}$CO$^+$(1--0) (or N$_2$H$^+$, see the label in the panels) emission lines toward the MDCs as seen with MOPRA. The lines are adjusted by a Gaussian (or with the \textit{hfs} procedure in CLASS to fit the multiplet of the N$_2$H$^+$) whose results are presented in Table~\ref{t:obs}. The frame on the bottom displays the residuals of the HCO$^+$(1-0) emission after subtraction of the Gaussian feature shown on the top frame.}
\label{f:ls-h1}
\end{figure*}

\begin{figure*}[h!]
\centerline{
\includegraphics[trim = 0cm 0cm 13cm 0cm , width=0.8\textwidth]{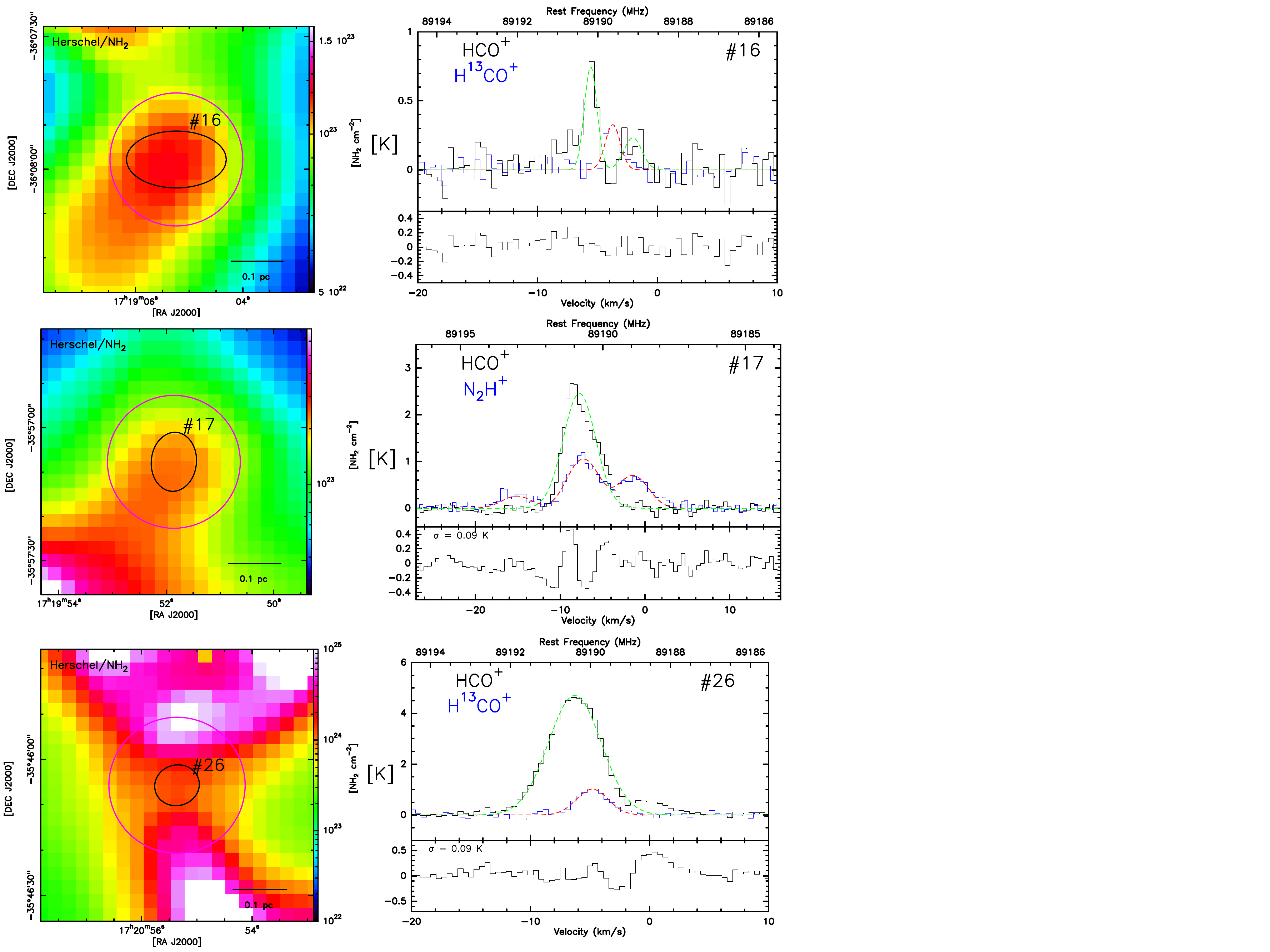}}
\caption{Same as Figure~\ref{f:ls-h1}, for the MDCs \#16, \#17, and \#26.}
\label{f:ls-h2}
\end{figure*}

\begin{figure*}[h!]
\centerline{
\includegraphics[trim = 0cm 0cm 13cm 0cm , width=0.8\textwidth]{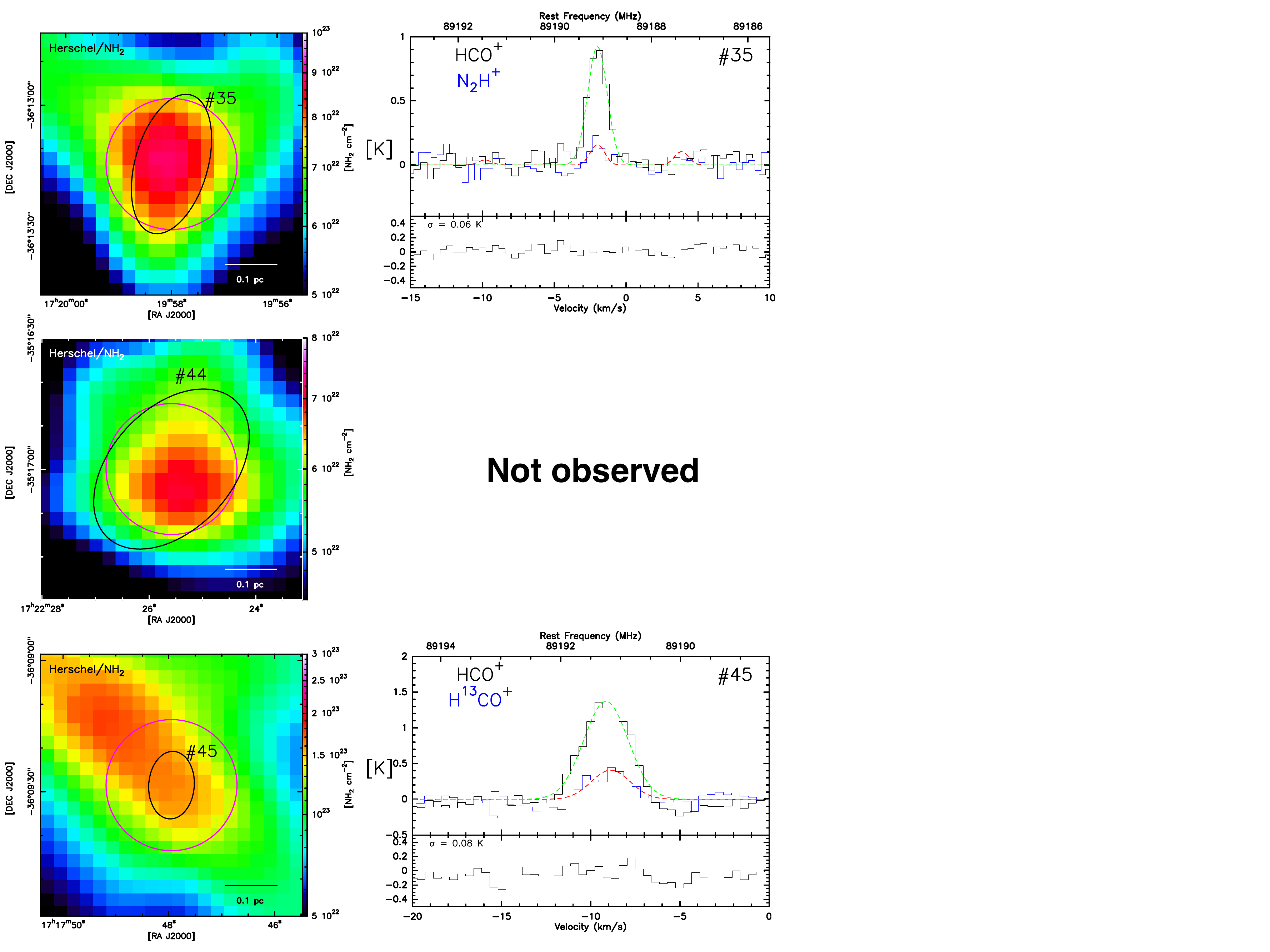}}
\caption{Same as Figure~\ref{f:ls-h1}, for the MDCs \#35, \#44, and \#45.}
\label{f:ls-h3}
\end{figure*}

Figure~\ref{f:ls} shows the \textit{Herschel}-250 $\mu$m emission of NGC6334, over-plotted with the positions of the 9 MDCs studied here. Figures~\ref{f:ls-h1}-\ref{f:ls-h2}-\ref{f:ls-h3} show the column density map over each MDC (left columns) and the molecular line emissions in the MDC (right columns). All observed MDCs are detected in HCO$^+$. Five MDCs are detected in H$^{13}$CO$^{+}$, and seven in N$_2$H$^+$ (see Table~\ref{t:obs}). Only the MDC \#11 was undetected in both N$_2$H$^+$ and H$^{13}$CO$^{+}$. We fitted the lines in \textit{CLASS}\footnote{see https://www.iram.fr/IRAMFR/GILDAS/doc/html/class-html/class.html.} using the \textit{hfs} procedure to fit the multiplet N$_2$H$^+$ molecular emission line and Gaussian functions to fit the HCO$^+$ and H$^{13}$CO$^{+}$ molecular line emissions (see Table~\ref{t:obs}). We used the fit to the N$_2$H$^+$ optically thin molecular line to determine the velocity dispersion in the MDCs as $\sigma_{\rm N_2H^+}= \dfrac{\rm FWHM_{\rm N_2H^+}}{\sqrt{8 \rm ln 2}}$. We used these velocity dispersion estimates $\sigma_{\rm N_2H^+}$ \footnote{Rigorously, the virial mass should be computed from an average molecule of weight $\mu\,m_p$ where $\mu=2.8$ \citep{kauffmann08} and $m_p$ is the mass of a proton. Considering the N$_2$H$^+$ as the molecule used to measure the dispersion, it gives  $\sigma=\sqrt{\sigma_{\rm N_2H^+}+\dfrac{k\,T}{m_p}\left(\dfrac{1}{2.8}-\dfrac{1}{29}\right)}$, which differs by less than 5\% with the value of $\sigma_{\rm N_2H^+}$.} to derive the virial mass for each MDC as:

\begin{equation}
M_{\rm vir}=3\left(\frac{5-2n}{3-n}\right)\times\frac{R\sigma_{\rm N_2H^+}}{G}
\end{equation}

where $R$ is the radius of the MDC, $G$ the gravitational constant, and $n$ is the index of the density profile $\rho(R)\propto R^{-n}$. \cite{garay07} and \cite{mueller02} find, on average, a radial profile index of 1.8 for massive dense cores. We use this value of $n=1.8$ to derive the virial mass of our MDCs.

We compute the virial parameter $\alpha_{vir}=M_{vir}/M_{MDC}$, where $M_{MDC}$ is the mass of the MDC, as reported by \cite{tige17}. The MDCs appear either close to virial equilibrium ($\alpha_{vir}\sim$0.8 for \#17, \#21, \#26, and \#45) or sub-virialized ($\alpha_{vir}\sim$0.35 for \#16, \#27, and \#35). This indicates that they are not transient objects and will most likely collapse and form stars.

\subsection{Infall motion in MDCs}
\label{ss:infall}

\begin{figure*}[ht!]
\centering
\includegraphics[width=1.0\textwidth]{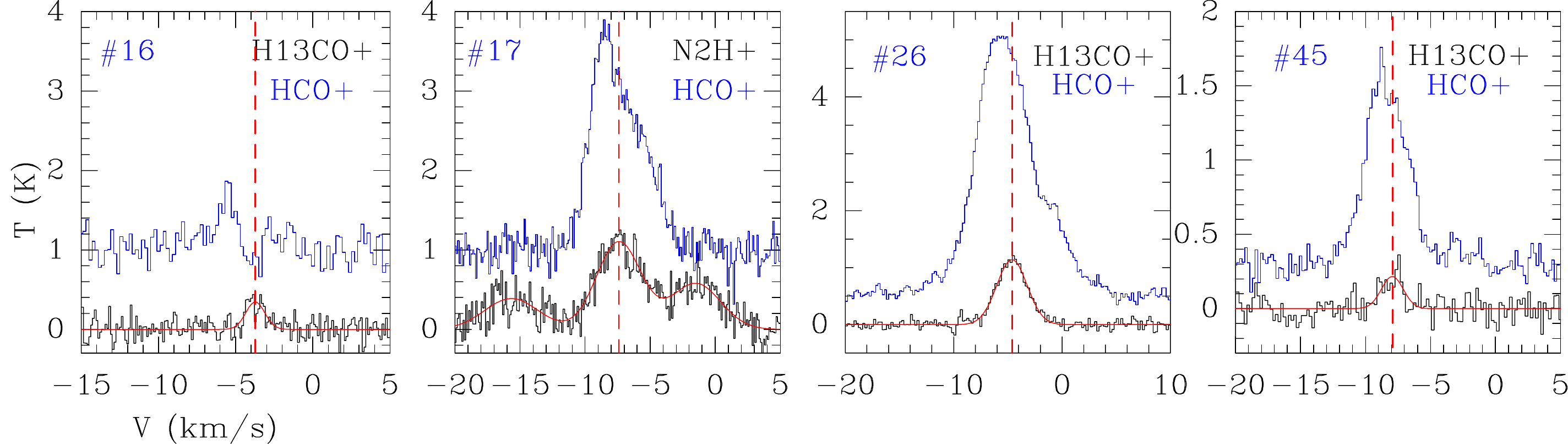}
\caption{Line profiles of the optically thick HCO$^{+}$(1-0) emission lines and of the optically thin H$^{13}$CO$^{+}$ -- respectively N$_2$H$^{+}$(1-0) -- emission lines towards the MDCs \#16, \#17, \#26 and \#45. The red dotted vertical line shows the velocity of the source determined from the optically thin line.}
\label{f:sud}
\end{figure*}

The majority of the MDCs show a blue-shifted central velocity of the optically thick HCO$^{+}$ emission line with respect to the optically thin H$^{13}$CO$^{+}$ and/or N$_2$H$^+$ emission lines (see Table~\ref{t:obs}). Following \cite{mardones97}, we quantify the asymmetry parameter  $\delta$V = (V$_{thick}$-V$_{thin}$)/$\sigma$V$_{thin}$ to measure line asymmetry, where V$_{thick}$ is the peak velocity of the optically thick line, V$_{thin}$ is the peak velocity of the optically thin line and $\sigma$V$_{thin}$ is the line width of the optically thin line. A negative $\delta$V value is indicative of infall, and a positive value of $\delta$V is indicative of expansion motions. The typical error in line asymmetry parameter is of 0.25, which is dominated by the error in determination of velocity of optically thick line. We report one positive value of $\delta$V (\#MDC 21) and 6 negative values of $\delta$V. The sources \#16, \#17, \#26 and \#45 further have $|\delta$V$|$ values above the typical error, which strongly indicate infall motions in these MDCs.

\cite{myers96} presented a simple model of contracting cloud, and showed that low-velocity contracting cloud display optically thick emission lines with double-peak spectra where the blue-shifted peaks appear stronger than the red-shifted peaks. At higher infall velocities the red-shifted peak of the optically thick emission lines disappear, displaying instead a shoulder-like profile. The four MDCs that have a significant negative $\delta$V value display such profiles. Figure~\ref{f:sud} presents the optically thick HCO$^{+}$ emission line superposed with the optically thin H$^{13}$CO$^{+}$ emission line (N$_{2}$H$^{+}$ for \#17) towards these 4 MDCs. In addition to the blue-shift of the optically thick line with respect to the optically thin line, the optically thick HCO$^{+}$ emission lines display a shoulder-like profile towards the MDCs \#17 and \#26, and a higher blue-peak with self absorption dips in the MDCs \#16, and \#45. Therefore, from the values of the $\delta$V parameters together with the profile of their optically thick lines, we conclude that these four MDCs are undergoing infalling motions.

\subsection{ALMA view of the massive dense cores}
\label{ss:ss}

\begin{figure*}[h!]
\centerline{
\includegraphics[trim = 0cm 1cm 6cm 0cm , width=1.0\textwidth]{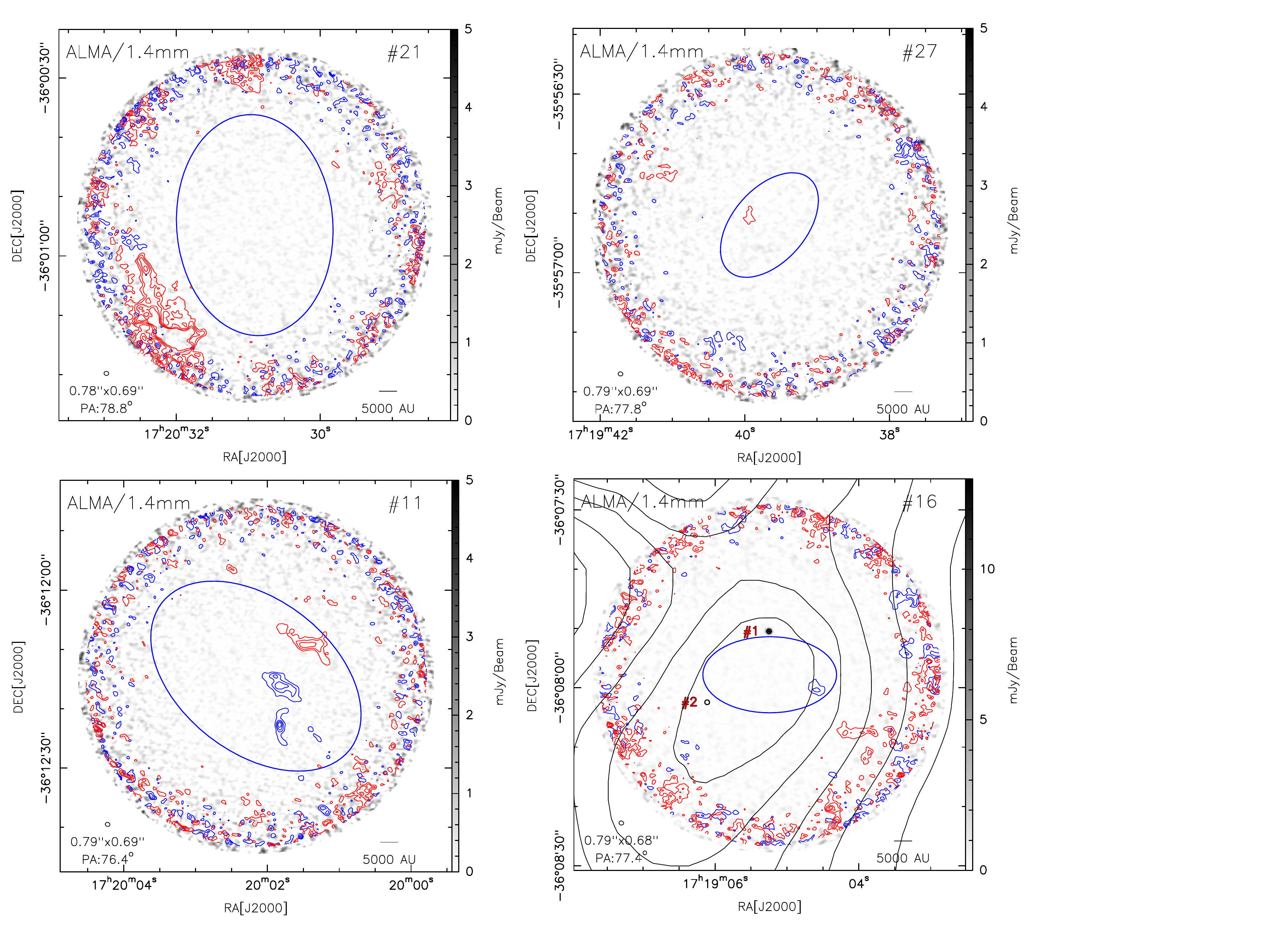}}
\caption{ALMA 1.4 mm continuum emission toward the 9 MDCs. The blue ellipses show the MDCs as defined by \cite{tige17}. The compact sources extracted by \textit{Getsources} in the MDCs \#16, \#26, and \#45 (see text and Table~\ref{t:alma} for details) are indicated with small black ellipses and red labels. The blue contours and red contours highlight the ALMA $^{12}$CO(2--1) integrated emission in the blue-shifted and red-shifted channels, respectively. The contours start at 3$\sigma$ with a 3$\sigma$ step in each panel. The values for the integration in velocities and the resulting rms is given in the Table~\ref{t:sup-alma}. In the bottom right panel (MDC \#16) the black contours show the column density emission derived from \textit{Hershel} observations \citep{tige17}. The contours range from 50\% to 90\% by step of 10\% of the peak flux in the MDC \#16 (blue ellipse). They show that the continuum sources ALMA\#1 and \#2 are associated with the MDC \#16.}
\label{f:alma1}
\end{figure*}

\begin{figure*}[h!]
\centerline{
\includegraphics[trim = 0cm 1cm 6cm 0cm , width=1.0\textwidth]{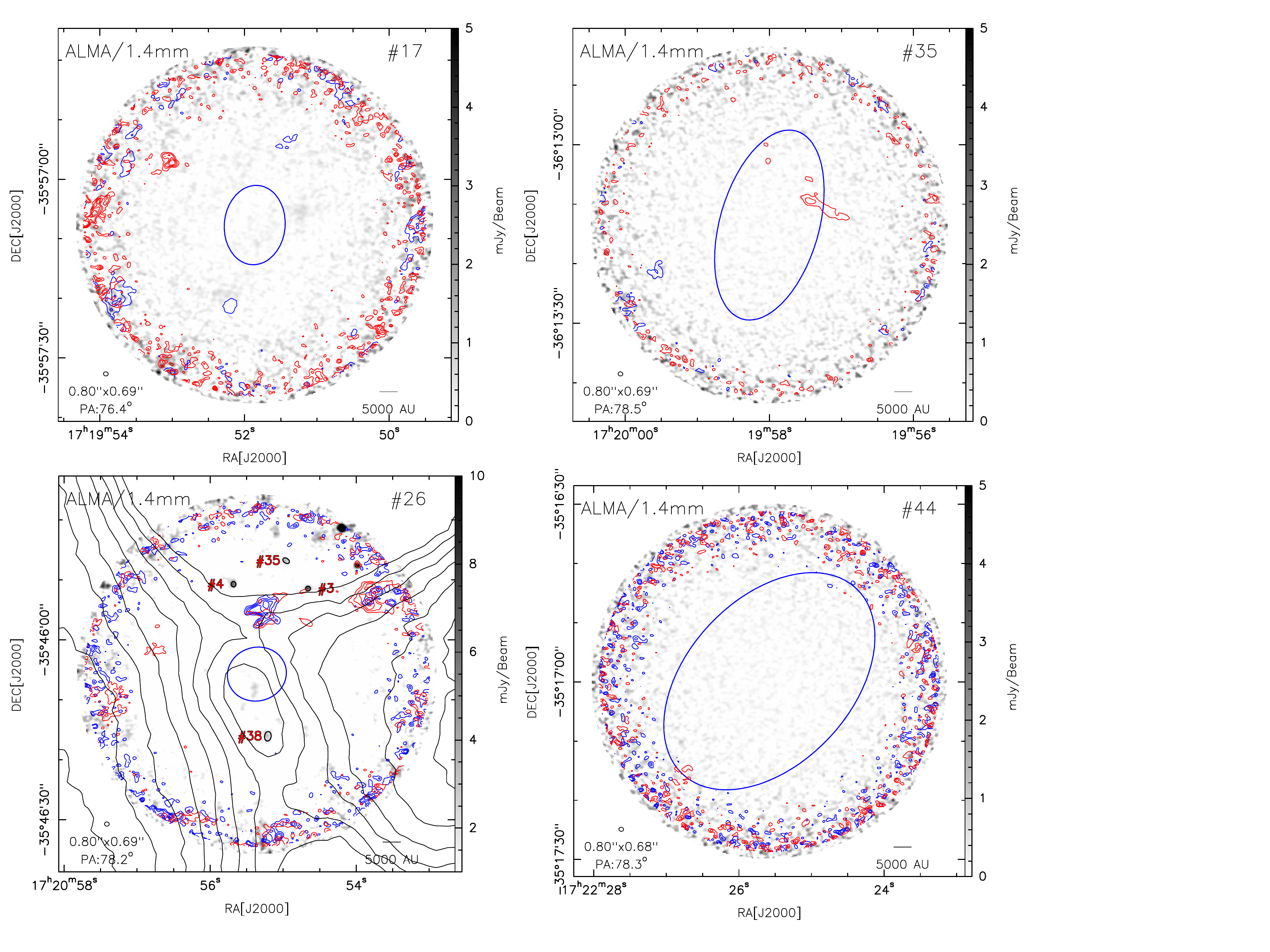}}
\caption{Following of Figure~\ref{f:alma1}. In the bottom left panel the black contours show the JCMT/SCUBA-2/450 $\mu$m emission presented in \cite{tige17}. The contours range from 20\% to 90\% of the peak flux inside the MDC \#26 (defined by the blue ellipse) by step of 10\%. They show that the continuum source ALMA\#38 is associated with the MDC \#26, while the continuum sources ALMA\#3, \#4, and \#35 belong to another structure.} 
\label{f:alma2}
\end{figure*}

\begin{figure}[h!]
\centerline{
\includegraphics[trim = 0cm 0cm 0cm 0cm , width=0.5\textwidth]{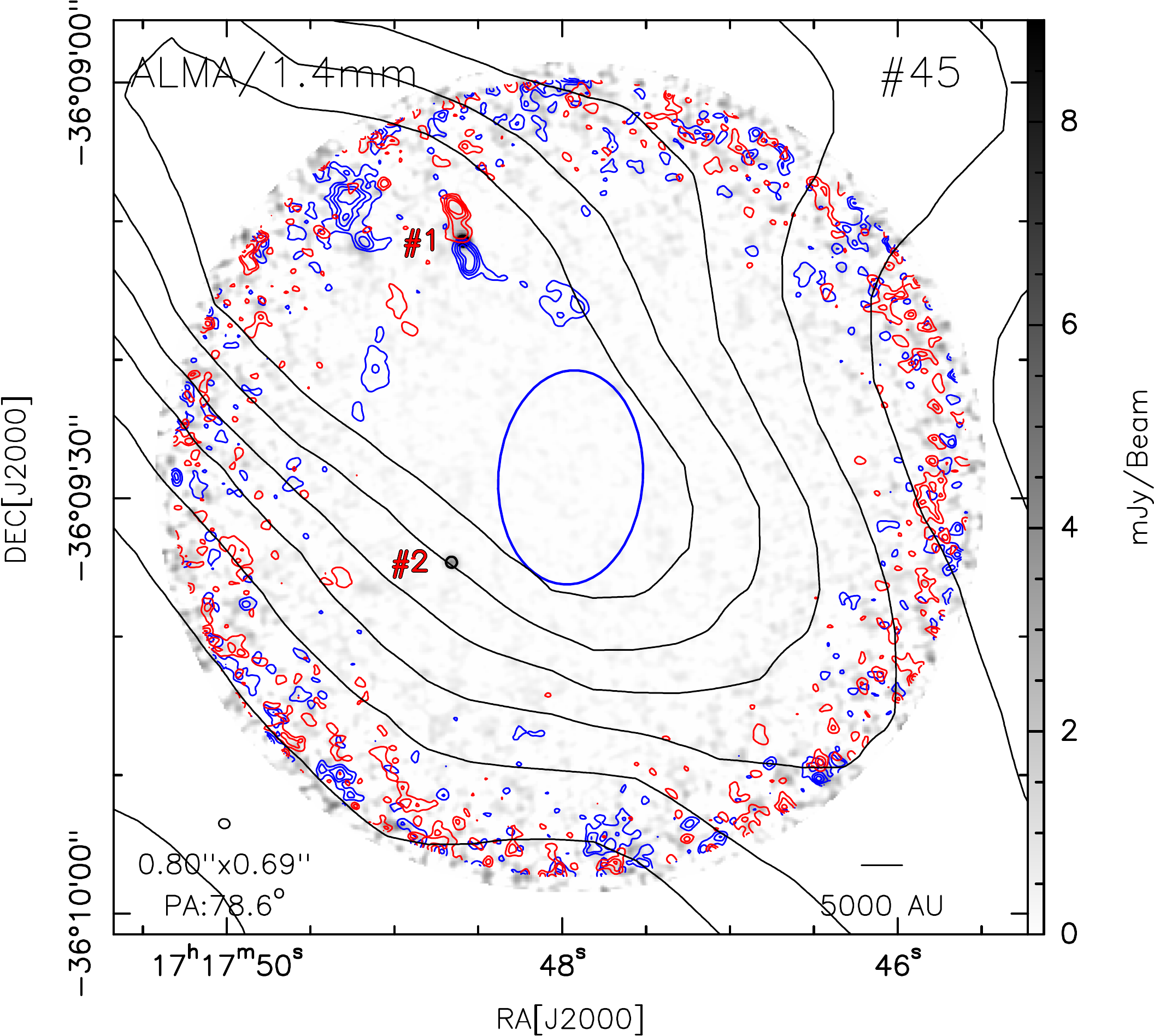}}
\caption{Following of Figure~\ref{f:alma2}. The black contours show the column density emission derived from \textit{Hershel} observations \citep{tige17}. The contours range from 30\% to 90\% of the peak flux in the MDC \#45 (blue ellipse) by step of 10\% . They show that the continuum sources ALMA\#1 and \#2 are not associated with the MDC \#45.}
\label{f:alma3}
\end{figure}

Figures~\ref{f:alma1}-\ref{f:alma2}-\ref{f:alma3} present  ALMA continuum 1.4mm emission maps toward the 9 starless MDCs. The emission is completely resolved out in the MDCs \#11, \#17, \#21, \#27, \#35, and \#44, indicating that they are emitting at a scale near or larger than our filtering scale of 11.6$^{\prime\prime}$ (or $\sim$0.1 pc at the adopted distance of NGC6334). On the contrary, the maps toward the MDCs \#16, \#26, and \#45 display some structure. To extract the compact sources within the fields we used the source extraction tool \textit{Getsources} \citep{sacha12}. Developed for multi-wavelength \textit{Herschel} images, it calculates the local noise and local background to properly extract compact sources from a complex cloud environment (see e.g. \citealt{louvet14}). In total we find 8 cores located in the vicinity of the MDCs \#16, \#26 and \#45. In these three fields, we superposed contours showing the column density emission derived from the \textit{Herschel} observations (respectively the JCMT/SCUBA-2/450$\mu$m emission for the MDC\#26). In the pointing toward the MDC \#16, two fragments are within the area corresponding to 90\% of the column density peak emission of the MDC. Therefore, these two fragments are likely forming from the mass reservoir of the MDC\#16. In a similar way, one fragment observed with ALMA seems associated with the MDC\#26. On the other hand, the 5 remaining fragments are offset from the MDC locations, and generally associated with proto-stellar objects previously identified by \cite{tige17}. The integrated flux of these cores varies between 1 mJy and 15 mJy. To derive the corresponding mass, we used the following expression:

\begin{equation}
M=\frac{S^{\rm int}_{\rm 1.4mm}\times d^2}{\kappa_{1.4mm}\times B_{1.4mm}}
\end{equation}

 with $d=1.75$ kpc, and $B_{\rm 1.4mm}$ the Planck function at the temperature $T_{\rm dust}$ given in Table~\ref{t:anex}. The dust mass opacity was taken to be equal to $\kappa_{1.4 mm} = 9.9 \times 10^{-3} cm^2 g^{-1}$, following the $\kappa_\nu = 0.1 cm^2 g^{-1} \times (\nu/1000 GHz)^\beta$ equation including a gas to dust ratio of 100 and an opacity index $\beta$ = 1.5, which is typical of dense and cool media \citep{ossenkopf94}. The mass of the cores ranges between 0.2 \msun~and 2.6 \msun~for a mean size of $\sim$0.008 pc. Compared with the mass of the MDCs ($\sim$110 \msun) it is clear that most of the emission is resolved out even in these MDCs where substructure is found. The parameters of each low-mass core are reported in Table~\ref{t:alma}.

\begin{table*}
\caption{Physical parameters of the fragments}
\label{t:alma}
\centerline{
\addtolength{\tabcolsep}{0pt}
\begin{tabular}{cccccc}
\hline
\hline  
Source$^a$       & RA          & DEC          & Integrated    & Size            & Mass$^b$      \\ 
                 & [J2000]     & [J2000]      & Flux [mJy]    & [''x'']         & \msun         \\ 
\hline                                                                                
MDC\#16-alma1       & 17:19:05.25 & -36:07:50.48 & 14.2$\pm$0.1  & 0.8$\times$0.8  & 2.63$\pm$0.02 \\
MDC\#16-alma2       & 17:19:06.11 & -36:08:02.44 &  1.0$\pm$0.1  & 0.8$\times$0.8  & 0.17$\pm$0.02 \\
MDC\#26-alma3       & 17:20:54.66 & -35:45:51:44 &  7.1$\pm$0.6  & 0.8$\times$0.8  & 0.45$\pm$0.04 \\ 
MDC\#26-alma4       & 17:20:55.68 & -35:45:50.72 &  8.8$\pm$0.5  & 0.9$\times$0.8  & 0.56$\pm$0.03 \\ 
MDC\#26-alma35      & 17:20:54.93 & -35:45:46.8  &  7.0$\pm$0.9  & 1.2$\times$0.8  & 0.44$\pm$0.05 \\
MDC\#26-alma38      & 17:20:55.21 & -35:46:16:07 & 14.4$\pm$0.7  & 1.6$\times$1.1  & 0.91$\pm$0.04 \\
MDC\#45-alma1       & 17:17:48.59 & -36:09:11.45 & 15.0$\pm$0.4  & 0.8$\times$0.8  & 1.53$\pm$0.04 \\   
MDC\#45-alma2       & 17:17:48.66 & -36:09:34.63 &  4.9$\pm$0.2  & 0.8$\times$0.8  & 0.50$\pm$0.02 \\    
\hline                                                                                                                                                                    
Mean                &             &              &  9.0$\pm$0.4  & 1.0$\times$0.8  & 0.90$\pm$0.03 \\ 
\hline           
\end{tabular}}
($^a$): The label of the source is defined as the name of the nearest MDC, plus the \textit{Getsources} extraction number. 
($^b$): The method to derive the mass follows eq.~2.
\end{table*}

The blue and red contours on Figs.~\ref{f:alma1}-\ref{f:alma2}-\ref{f:alma3} show the blue-shifted and red-shifted integrated emissions of the $^{12}$CO(2-1), highlighting the presence of outflows. The three compact sources \#16-alma1, \#16-alma2, \#26-alma38 are deprived of outflows and are therefore pre-stellar core candidates. We report the detection of a bipolar outflow associated with the proto-stellar object \#45-alma1. 
We also report two $^{12}$CO blue-shifted blobs, plus one $^{12}$CO red-shifted blob in the MDC \#11;  and one elongated $^{12}$CO red-shifted emission in the MDC \#35. These $^{12}$CO emissions are not associated with any compact continuum sources, down to our detection limit of 0.11 \msun~ corresponding to a 5$\sigma$ level. The interpretation of these features goes beyond the scope of the present paper and will require the addition of complementary data, e.g. from ACA, to recover the intermediate scale emission.

\section{Discussion} 
\label{s:discu}

\subsection{Comparison with models}
\label{ss:models}

\cite{baldeschi17} find an empirical high-mass star formation threshold, based on studies of environment with and without high-mass star formation. They suggest that cloud structures with masses larger than the mass $m_{threshold}=1280$ \msun $(r/pc)^{1.42}$, where $r$ is the source radius, form high-mass stars or will form high-mass stars in the near future. Applying this relationship to our $\sim$0.1 pc MDCs, we find that all have masses above the mass threshold for high-mass star formation (see Table~\ref{t:obs}-column 11).

Our 9 MDCs have a mean mass of $\sim$115 \msun~and a mean radius of 0.1 pc. These conditions are very similar with the initial conditions of the quasi-static models for high-mass star formation \citep[100-200 \msun~and radius of 0.1 pc, see e.g.][]{mckee-tan02,krumholz07}. Three of the starless MDCs are clearly subvirialized with $\alpha_{vir}\sim$0.35, and three additional MDCs display evidence of infall from their large-scale molecular line emissions, in contrast with the virial equilibrium assumption used in these models. In the framework of quasi-static models, where the MDCs ($r\sim$0.1 pc) slowly contract down to the physical scale at which individual (or low N-tuple system) star formation takes place \citep[$\sim$0.03 pc, see e.g. ][]{bontemps10, zhang14b, palau15}, one expects to observe high-mass pre-stellar cores. Nevertheless, the ALMA observations toward NGC6334 showed that the few cores that were detected are low-mass cores (maximum mass of 2.6 \msun), and hence they do not have the mass reservoir to form a high-mass star.

On the other hand the competitive accretion models \citep[e.g.][]{bonnell04} posits that an MDC of mass $M_{\rm MDC}$ will fragment into $n$ cores where each core will have a mass close to the thermal Jeans mass ($M_{\rm Jeans}$), and $n\simeq\dfrac{M_{\rm MDC}}{2.5\times M_{\rm Jeans}}$. From the local cloud density and kinetic temperature computed for each MDCs (see Table~\ref{t:obs} and \citealt{tige17}), we can calculate the Jeans mass and Jeans scale ($\lambda_{\rm Jeans}$) as:

\begin{align}
M_{\rm Jeans}&=\frac{\pi^{5/2}}{6}\frac{\sigma_{\rm th}^3}{\sqrt{\rm G^3 \rho}},~ ~ ~ ~ ~ \lambda_{\rm Jeans}=\sigma_{\rm th}\sqrt{\frac{\pi}{\rm G \rho}} 
\end{align}

where G is the gravitational constant, and $\rho$ is the gas density (given in Table~\ref{t:anex}-column 6).  $\sigma_{\rm th}$ is the thermal broadening of the emission line given by $\sigma_{th}=\sqrt{k_B\times T/(\mu\times m_p})$, where $k_B$ is the Boltzman constant, $\mu=2.8$ the mean molecular weight \citep{kauffmann08}, $m_p$ the mass of a proton, and $T$ is the temperature of the MDC. We obtained a mean Jeans mass $<$M$_{Jeans}>$=0.65~\msun~and a mean Jeans length of $<\lambda_{Jeans}>$=0.03~pc. Our ALMA data probe angular scales down to 0.74$^{\prime\prime}$ ($\sim$0.006 pc at the distance of NGC6334), and have a mass equivalent 5$\sigma$ detection threshold of 0.11 \msun, hence we are fully sensitive to the Jeans mass and Jeans scale. Therefore if a cluster of low-mass cores were forming in our MDCs we should have detected it. In the case of the MDC \#16 where we found cores associated with the mass reservoir of the MDC, a fragmentation such as predicted by the competitive accretion models would have resulted in $\sim$280 cores of $\sim$0.2 \msun~in mass. In contrast we only found two cores of masses of 0.2 \msun~and 2.6 \msun, respectively. In the MDC \#26 a fragmentation such as in the competitive accretion models would have resulted in $\sim$120 cores of $\sim$0.3 \msun~in mass, whilst we found only a single core of $\sim$0.9~\msun~in mass. In the remaining 7 MDCs no core is found associated with the starless MDCs. Therefore the observations of starless MDCs in NGC6334 do not match the predictions of competitive accretion models, unless the MDCs are in such early stage of evolution that the initial fragmentation has not yet taken place. Most of the mass in the starless MDCs is likely organized in extended and homogeneous cloud structures ($\ga$0.1~pc) that were filtered by the interferometer. A similar result was reported by \cite{dunham16} who reported that low-mass pre-stellar cores resembling Bonnort-Ebert spheres would be filtered out by the interferometers. Combination with short-spacing observations is therefore mandatory to investigate the exact cloud structure of starless MDCs.

It is not clear what prevents the MDCs from fragmenting into low-mass cores. According to \cite{krumholz-mckee08} the heat produced by accreting low-mass stars in regions with surface density of at least 1 g cm$^{-2}$ can halt fragmentation of the MDCs. It could be the case in MDCs \#11 and \#35 if the $^{12}$CO(2-1) blobs (see Sect.~\ref{ss:ss}) were interpreted as monopolar outflows. The combination of the heat due to the accretion on the low-mass protostars with the injection of turbulence by the outflows could prevent the fragmentation in the MDCs. But the lack of fragmentation is also observed in the MDCs where no sign of proto-stellar activity is present (MDCs \#16-17-21-26-27-44-45, see Figs.~{\ref{f:alma1}-\ref{f:alma2}-\ref{f:alma3}). Consequently further processes must be considered.

Numerical models of a protoclusters where a significant magnetic field is considered show a smaller fragmentation level than models with weak magnetic support \citep[e.g.,][see also the theoretical approach by \citealt{shu87}]{commercon11}. Therefore, the magnetic field could play the dual role of \textit{(i)} limiting the fragmentation of MDCs and \textit{(ii)} sustaining the MDCs against collapse together with the turbulence. Following this hypothesis, we derive the magnetic field strength necessary to prevent the MDCs from collapsing given their measured turbulence support. The magnetic field strength is computed as $B=\sigma_A\times\sqrt{\mu_0\,\rho}$, where $\rho$ is the mean density in the MDCs (see Table~\ref{t:anex}), $\mu_0$ is the vacuum permeability and $\sigma_A$ is the Alfvén velocity such as  M$_{\rm vir, B}$ = M$_{\rm MDC}$ = $3\frac{R}{G}\left(\frac{5-2n}{3-n}\right)\left(\sigma^2+\frac{1}{6}\sigma_A^2\right)$, where the parameters are the same as for eq~1. The value of the magnetic field necessary to sustain the MDCs against collapse is given in Table~{\ref{t:obs}}. The mean value of $\sim$1.2 mG that we obtain is of the order of the magnetic field strength measured in high-mass star formation \citep[$\sim$0.1-10 mG, see ][]{mbl18}. It is very likely that the magnetic field plays a central role in our MDCs. This is particularly true toward the MDC \#27, and \#35 where the virial parameter $\alpha_{\rm vir}<0.4$ whereas no infall motion are traced by the molecular line emissions. This discrepancy is better explained if magnetic fields are taken into account.

\subsection{Comparison with previous observational studies}
\label{ss:dobs}

Recently, a handset of large-sample observational studies have targeted high-mass star-forming regions with interferometers \citep{sanhueza17, kong17, csengeri17}. First, it is important to note that none of these observations found a thermal Jeans fragmentation and none could claim the discovery of a high-mass pre-stellar core -- in a similar way as the present study. Interestingly, these studies found different values in term of fragmentation level and/or on the masses of the cores. Here, it is primordial to remember that these studies probe different physical scales, with different filtering scales. For instance, \cite{csengeri17} report ALMA 7m array observations that probe scales from $\sim$0.3 pc down to $\sim$0.06 pc in 35 IR Quiet clumps selected from the ATLASGAL survey \cite{schuller09}. At these scales they find that the clumps fragment on average in 3 MDCs of typically 60 \msun, similar to the sample we scrutinized here with ALMA, down to scales of $\sim$0.006 pc. \cite{csengeri17} report that they were able to recover between 16\% and 47\% of the large scale fluxes of the clumps measured by the ATLASGAL survey at a similar wavelength of 870 $\mu$m. It is consistent with our observations that report higher filtering when probing the MDCs at higher angular resolution.

Also we stress that all these studies were conducted in different regions, hence scrutinizing different stages of the high-mass starless phase. Looking in details at these studies, they actually seem to fall well onto the empirical sequence proposed by \cite{mbl18}. Following this sequence, \textit{(i)} a starless MDC ($\sim$100 \msun~within 0.1 pc) would fragment into a few pre-stellar objects of low/intermediate masses, as we observed here in the MDCs \#16 and \#26. This type of fragmentation was also observed by \cite{kong17}: they surveyed 32 high-mass surface density regions in which they detected in total 141 cores, stating that many of the weakest sources are likely to be noise fluctuation. They give the characteristics of the 50 more reliable extractions and present in details the 6 most massive. The 6 most massive have a mean mass of 3.8 \msun~ if we don't consider the most massive (C9A, see below). Therefore, all the 141 cores of \cite{kong17} but C9A are low or intermediate mass cores. The most massive C9A core has a mass of $\sim$70 \msun~under their assumptions, and of $\sim$20 \msun~using our mean temperature of 15.5 K and our dust emissivity assumption. Also, it could be constituted of two lower mass cores, and, its evolutionary status is unknown. \textit{(ii)} Then, in the sequence of \cite{mbl18} the global collapse of the MDC will generate gas flows streams that will increase the mass of the pre-stellar objects - a process even more efficient as the pre-stellar objects are located near the gravitational potential of the MDC. This step is well supported by the observations by \cite{sanhueza17}. They observe a candidate starless clump (1500 \msun~within 0.6 pc) that fragments into 5 candidate pre-stellar cores with masses ranging from 8.4 \msun~to 15 \msun within $\sim$0.05 pc. It is also supported by the present study since the two MDCs \#16 and \#26, that harbour substructure, are collapsing (see Sect.~\ref{ss:infall}). \textit{(iii)} During the accretion process the low/intermediate mass pre-stellar cores eventually start proto-stellar activities, which would explain why some IR Quiet MDCs  host low/intermediate protostars (Louvet et al. in prep). \textit{(iv)} Finally, the low/intermediate protostars eventually become massive ($\sim$25 \msun~within 0.03 pc) and form high-mass stars.

To strengthen this view of high-mass star formation, future studies should target early-type proto-stellar MDCs ($\sim$100 \msun~within $\sim$0.1 pc) to determine the mean mass at which 0.03 pc cores turn proto-stellar.

\section{Conclusions}
\label{s:conclu}

We presented 1.4 mm continuum and $^{12}$CO(2--1) ALMA observations of 9 starless massive dense cores (MDCs) located in NGC\,6334 ($d\sim$1.75 kpc) that have a mean mass of 115 \msun~and a radius of $\sim$0.1 pc. Our ALMA observations probe spatial scales from $\sim$0.1 pc down to 0.006 pc, resolving the thermal Jeans length in these 9 MDCs ($<\lambda_{\rm Jeans}>$= 0.03 pc). Our ALMA observations are also sensitive to the thermal Jeans mass in the MDCs ($<M_{\rm Jeans}>$=0.65 \msun), with a mass equivalent 5$\sigma$ level of $\sim$0.1 \msun. We complemented the ALMA data-set with archive HCO$^+$(1--0), H$^{13}$CO$^+$(1--0), and N$_2$H$^+$(1--0) MOPRA observations toward 8 of the 9 MDCs to examine the gas dynamic at the MDCs' scale. Our main results can be summarized as follows:
\begin{itemize}
\item[$\bullet$] The 9 starless MDCs of our sample have the potential to form a high mass star according to the mass threshold criteria presented by \cite{baldeschi17}.
\item[$\bullet$] Three MDCs are subvirialized with $\alpha_{\rm vir}\sim0.35$ and four MDCs are roughly virialized with $\alpha_{\rm vir}\sim 0.8$. One among the former and three among the later MDCs further show an asymmetry of their optically thick HCO$^+$ molecular line emissions and blue-shifted emission lines with respect to the optically thin H$^{13}$CO$^+$ emission lines, indicative of collapse.
\item[$\bullet$] ALMA observations show that very few 0.005 pc cores are forming in starless MDCs. In total, three cores are detected: two in the MDC \#16 with masses of $\sim$0.2 \msun~and 2.6 \msun~respectively, and one in the MDC \#26 with a mass of 0.9 \msun. These three cores are deprived of outflows making them candidate pre-stellar cores. These low-mass dense cores do not have the gas reservoir to form a high-mass star. Therefore most of the gas must be organized into extended and homogeneous structures near or above our filtering scale ($\ga$0.1~pc).
\item[$\bullet$] These ALMA observations do not support the turbulent-core models of high-mass star formation since we do not observe high-mass pre-stellar cores (expected to have $\sim$25 \msun~within $\sim$0.03 pc).
\item[$\bullet$] These ALMA observations do not reproduce the predictions of the competitive accretion models in term of initial fragmentation of the MDCs. We detected only a few fragments in our sample, two orders of magnitude below the predicted fragmentation level. Our observations are therefore incompatible with these models unless the initial fragmentation has not yet taken place.
\item[$\bullet$] Present ALMA observations of the starless MDCs of NGC6334 are consistent with the empirical model by \cite{mbl18} in which the first stage of the high-mass star formation scenario is the formation of a few low/intermediate mass pre-stellar cores within the MDCs, that will grow in mass by accreting gas from the diffuse gas reservoir of the MDC.
\end{itemize}

\begin{acknowledgements}
\noindent{The authors thank the anonymous referee  for a thorough report which led to a significant improvement of this paper.}
\noindent{FL acknowledges the help of Arnaud Belloche in the interpretation of optically thick HCO$^+$ emission lines. FL thanks Timea Csengeri for useful discussions about the main results of the article.}
\noindent{FL acknowledges the support of the Fondecyt program nº 3170360. }
\noindent{FL acknowledges the support of the Programme National de Physique Stellaire and Physique et Chimie du Milieu Interstellaire of CNRS/INSU (with INC/INP/IN2P3), co-funded by CEA and CNES.}
\noindent{G.G. and L.B. acknowledges support by CONICYT project Basal AFB-170002.}
\noindent{This paper makes use of the following ALMA data: ADS/JAO.ALMA\#2015.1.00850.S. ALMA is a partnership of ESO (representing its member states), NSF (USA) and NINS (Japan), together with NRC (Canada), NSC and ASIAA (Taiwan), and KASI (Republic of Korea), in cooperation with the Republic of Chile. The Joint ALMA Observatory is operated by ESO, AUI/NRAO and NAOJ.}
\end{acknowledgements}

\bibliographystyle{aa} 
\bibliography{fab} 

\section{Appendix}

\begin{table*}
\caption{Additional parameters of the MDCs}
\label{t:anex}
\centerline{
\addtolength{\tabcolsep}{+1pt}
\begin{tabular}{lcccccc|ccc}
\hline
\hline 
                                    \multicolumn{7}{c}{Data-set from \cite{tige17}}                                                           &          \multicolumn{3}{c}{ALMA-cycle 3 observations}               \\
\hline
Source$^a$               & RA                & DEC            & L$_{bol}^a$   & M/L                    & $<n_{\rm H_2}>^a$   & T$^a$   & $\sigma_{^{12}CO}^b$   & $\sigma_{cont}$ & Mass Sensitivity                 \\
                         & [J2000]           & [J2000]        & [L$_\odot$]   & [M$_\odot$/L$_\odot$] & [cm$^{-3}$]          & [K]     & [mJy/Beam]             & [$\mu$Jy/Beam]  & at 5$\sigma^c$                   \\
\hline                                                                                                                                                                  
NGC6334-MDC\#11          & 17:20:02.16       & -36:12:14.5    & 110           &  1.55                  & 2.4$\times10^5$     & 14.1    &  15.8                  & 273             & 0.11 \msun                       \\
NGC6334-MDC\#16          & 17:19:05.24       & -36:07:57.8    &  11           & 13.64                  & 1.6$\times10^6$     &  9.9    &  15.8                  & 263             & 0.19 \msun                       \\
NGC6334-MDC\#17          & 17:19:51.86       & -35:57:07.7    & 230           &  0.65                  & 4.8$\times10^6$     & 16.5    &  16.8                  & 310             & 0.10 \msun                       \\
NGC6334-MDC\#21          & 17:20:30.91       & -36:00:54.8    & 390           &  0.33                  & 2.2$\times10^5$     & 18.3    &  16.9                  & 283             & 0.06 \msun                       \\
NGC6334-MDC\#27          & 17:19:39.66       & -35:56:52.0    & 120           &  0.88                  & 1.3$\times10^6$     & 15.6    &  16.4                  & 292             & 0.07 \msun                       \\
NGC6334-MDC\#26          & 17:20:55.36       & -35:46:05.7    & 740           &  0.14                  & 7.5$\times10^6$     & 21.1    &  17.6                  & 656             & 0.16 \msun                       \\
NGC6334-MDC\#35          & 17:19:58.00       & -36:13:13.5    & 140           &  0.64                  & 3.4$\times10^5$     & 16.4    &  17.2                  & 318             & 0.11 \msun                       \\
NGC6334-MDC\#44          & 17:22:25.58       & -35:16:59.9    &  27           &  2.85                  & 1.0$\times10^5$     & 12.8    &  16.5                  & 281             & 0.08 \msun                       \\
NGC6334-MDC\#45          & 17:17:47.95       & -36:09:28.5    &  60           &  1.27                  & 2.0$\times10^6$     & 14.7    &  16.5                  & 352             & 0.14 \msun                       \\   
\hline                                                                                                                                                                                                                    
mean                     & -                 & -              & 203           &  2.44                  & 2.0$\times10^6$     & 15.5    &  16.6                  & 336             & 0.11 \msun                       \\     
\hline                                                                                                                                                                                                                                                                                                                                    
\end{tabular}}
\textbf{(a)}: The sources name, radius, bolometric luminosity, and density follow the work by \cite{tige17}.    
\textbf{(b)}: The 1$\sigma$ rms noise level is given per channel, where the channel width is of 0.7 \kms.
\textbf{(c)}: The method to estimate the mass from the 5$\sigma$ value follows the equation (2).
\end{table*}

\begin{table*}
\caption{Integration parameters of the CO line in the MDCs}
\label{t:sup-alma}
\centerline{
\addtolength{\tabcolsep}{+1pt}
\begin{tabular}{l|cc|cc}
\hline
\hline                                                                                                                                                                                  
MDC                   &      \multicolumn{2}{c}{Blue-shifted $^{12}$CO(2--1) emission}  &    \multicolumn{2}{c}{Red-shifted $^{12}$CO(2--1) emission}   \\
\hline
                      & Velocity range              & 1$\sigma$ value                   &        Velocity range        & 1$\sigma$ value                \\
                      & [km/s]                      & mJy/beam.\kms                     &        [km/s]                & mJy/beam.\kms                  \\
\hline                                                                                                                                                         
NGC6334-MDC\#11       & [-19.4 ; -7.9 ]             & 11.1                              & [0.0 ; 3.1 ]                 & 6.2                            \\   
NGC6334-MDC\#16       & [-8.5 ; -5.5 ]              & 11.9                              & [1.0 ; 6.1 ]                 & 7.3                            \\   
NGC6334-MDC\#17       & [-13.3 ; -8.9 ]             & 20.1                              & [-0.1 ; 5.7 ]                & 5.7                            \\    
NGC6334-MDC\#21       & [-12.6 ; -9.7 ]             & 4.5                               & [10.9 ; 14.2 ]               & 7.9                            \\     
NGC6334-MDC\#27       & [-18.6 ; -9.3 ]             & 8.2                               & [-2.2 ; 6.4 ]                & 9.3                            \\     
NGC6334-MDC\#26       & [-30.1 ; -20.5 ]            & 7.8                               & [-7.9 ; -5.3 ]               & 15.9                           \\    
NGC6334-MDC\#35       & [-9.8 ; -4.5 ]              & 7.4                               & [1.8 ; 5.4 ]                 & 4.8                            \\      
NGC6334-MDC\#44       & [-16.5 ; -7.5 ]             & 6.3                               & [1.5 ; 10.7 ]                & 6.6                            \\      
NGC6334-MDC\#45       & [-23.3 ; -17.0 ]            & 9.6                               & [0.5 ; 13.3 ]                & 9.9                            \\  
\hline    
\end{tabular}}
\end{table*}

\end{document}